\documentclass[a4paper,12pt]{article}
\usepackage[]{epsfig}
\usepackage[]{rotating}
\usepackage[]{cite}

\newcommand{\ee}{\mbox{${\mathrm{e}}^+ {\mathrm{e}}^-$}}

\newcommand{\qq}         {\mbox{$\mathrm{q}\bar{\mathrm{q}}$}}

\newcommand{\lpair}      {\mbox{$\ell^+\ell^-$}}
\newcommand{\nunu}       {\mbox{$\nu\bar{\nu}$}}

\newcommand {\Ho}        {\mbox{$\mathrm{H}^{0}$}}

\newcommand {\ho}        {\mbox{$\mathrm{h}^{0}$}}
\newcommand {\Zo}        {\mbox{$\mathrm{Z}^{0}$}}

\newcommand{\gaga}       {\mbox{$\gamma\gamma$}}

\def\mrm       {\mathrm}

\newcommand{\sqrts}     {\sqrt{s}}
\newcommand{\PhysLettB}[1] {Phys. Lett. {\bf B#1}}
\newcommand{\CPC}[1]      {Comp.\ Phys.\ Comm.\ {\bf #1}}
\def\NIM                {\mbox{Nucl. Instr. Meth.}}

\newcommand{\ra}        {\mbox{$\rightarrow$}}   
\begin{document}

\begin{titlepage}
\centerline{\Large EUROPEAN ORGANIZATION FOR NUCLEAR RESEARCH}

\begin{flushright}
      LHWG Note 2001-08 \\
      ALEPH 2001-059 CONF 2001-039 \\
      DELPHI 2001-116 CONF 539 \\
      L3 Note 2701 \\
      OPAL Technical Note TN695 \\
      July 2, 2001 \\   
\end{flushright}

\bigskip
\bigskip
\bigskip
\bigskip
\bigskip

\begin{center}{\LARGE\bf Searches for Higgs Bosons Decaying into
Photons: Preliminary Combined Results Using LEP Data Collected at
Energies up to 209~GeV}
\end{center}
\bigskip
\begin{center}{\LARGE The ALEPH, DELPHI, L3 and OPAL Collaborations\\
The LEP Higgs Working Group}
\end{center}


%
%

\bigskip\bigskip

\bigskip
\begin{center}{\large  Abstract}\end{center}
This note describes preliminary LEP-combined results of searches 
for Higgs bosons decaying into photons. 
The analyses use data collected at 
$\sqrt{s} \approx $88$-$209~GeV.
Using the combined data for \Ho\ra\gaga,
a lower bound of 108.2~GeV is set at the 95\% confidence level 
for the ``benchmark'' fermiophobic Higgs boson

\bigskip
\bigskip
%
\begin{center}
  {\Large
   THE RESULTS QUOTED IN THIS NOTE ARE PRELIMINARY}
\end{center}

\end{titlepage}

%
%
\newpage
\section{Introduction}

In the Standard Model (SM), the branching fraction
for \ho\ra\gaga\
is too small to permit its observation at LEP energies.
However, in particular formulations of 
2-Higgs Doublet Models (2HDM)
and other models,
the Higgs coupling to fermions can be small
and the Higgs bosons therefore decay preferentially to pairs of bosons.
These are the so-called ``fermiophobic'' Higgs bosons
(see, e.g., references~\cite{Akeroyd,Santos}).

The fermiophobic models are indeed parameter dependent,
but a large class of the models has near-Standard Model production
strength.
Therefore, we establish a ``benchmark'' fermiophobic model
defined to be the Standard Model production rates and channels,
but with the fermionic channels closed.
To model Higgs boson production 
the four LEP experiments are using the code 
HZHA~\cite{HZHA3} and/or HDECAY~\cite{HDECAY}. 
The code 
HDECAY gives a slightly lower di-photon branching fraction than HZHA
(the difference in mass limits near 100~GeV is about 0.6~GeV).  
In the interest of conservatism, we use
the HDECAY branching fractions in this LEP-wide combination.
The bosonic branching fractions obtained from HDECAY in the benchmark
model are shown in Figure~\ref{hgg-br}.

To obtain the limits, the LEP Higgs Working Group uses the frequentist confidence levels
generated by the ``ecl'' package~\cite{ecl}; three of the LEP
collaborations use different statistical methods, which have been
shown to agree with the LHWG method to within 0.5~GeV on all mass limits. This
technique includes the estimated systematic errors and the expected signal shape in
setting test-mass dependent limits on Higgs boson decays to photon pairs.

The LEP limits on the fermiophobic Higgs boson mass will be shown to
exceed 100~GeV, where Figure~\ref{hgg-br} indicates preferred decay
into WW.  One LEP experiment~\cite{L3WW} has set limits on the WW
mode, but this channel is not considered in the present combination.

%
\section{\boldmath Searches for \ho\Zo\ with \ho\ra\gaga }

The 
ALEPH~\cite{ALEPH}, 
DELPHI~\cite{DELPHI}, 
L3~\cite{L3}, and 
OPAL~\cite{OPAL}
analyses are described in journal
articles or CERN preprints.
All of the LEP experiments search for hadronic, leptonic, and missing
energy (neutrino) decay modes of the associated Z boson in the
production channel \ee \ra \ho\Zo.  
The ALEPH experiment does not
discriminate between Z decay modes; rather, it performs a ``global''
analysis which focusses on identifying the di-photon state.  DELPHI,
L3, and OPAL seek to identify the 
hadronic (\Zo\ra\qq), 
leptonic (\Zo\ra\lpair), 
and 
missing energy (\Zo\ra\nunu) 
classes. 

The LEP data from energies just below the Z resonance to the highest
LEP-2 energy of 209 GeV can be used for this search.  
The energies and Z decay modes used by the 4 LEP experiments are
summarized in Table~\ref{T:hgg1}.

%
%
\begin{table}[htbp]
\begin{center}
{\footnotesize
\begin{tabular}{|l||c|c|c|c||c|}\hline
        & ALEPH & DELPHI & L3  & OPAL & SUM \\ 
\hline\hline

Modes     & global & \qq,\lpair,\nunu\ & \qq,\lpair,\nunu\ & \qq,\lpair,\nunu\ & \\
$\sqrt{s}$ (GeV) & 192-209 & 189-209 & 189-209 & 88-209 & 88-209 \\
Candidates      & 10   & 47   & 64   & 184   & 305 \\
Background      & 10.8 & 42.2 & 69.5 & 185.7 & 308.2\\
Benchmark limit (GeV) & 104.4 & 103.6 & 104.1 & 104.8 & 108.2 \\
Expected limit (GeV)  & 104.6 & 105.1 & 104.9 & 105.2 & 109.0 \\
\hline
\end{tabular}
}
\end{center}
  \caption[Datasets and Limits]
  {
    Parameters of the photonically-decaying Higgs searches.
95\% confidence level lower limits on the mass of a benchmark
  fermiophobic Higgs boson are shown.
}
  \label{T:hgg1}
\end{table}

\begin{figure}[!htb]
        \vspace{0.8cm}
        \begin{center}
           \resizebox{\linewidth}{!}{\includegraphics{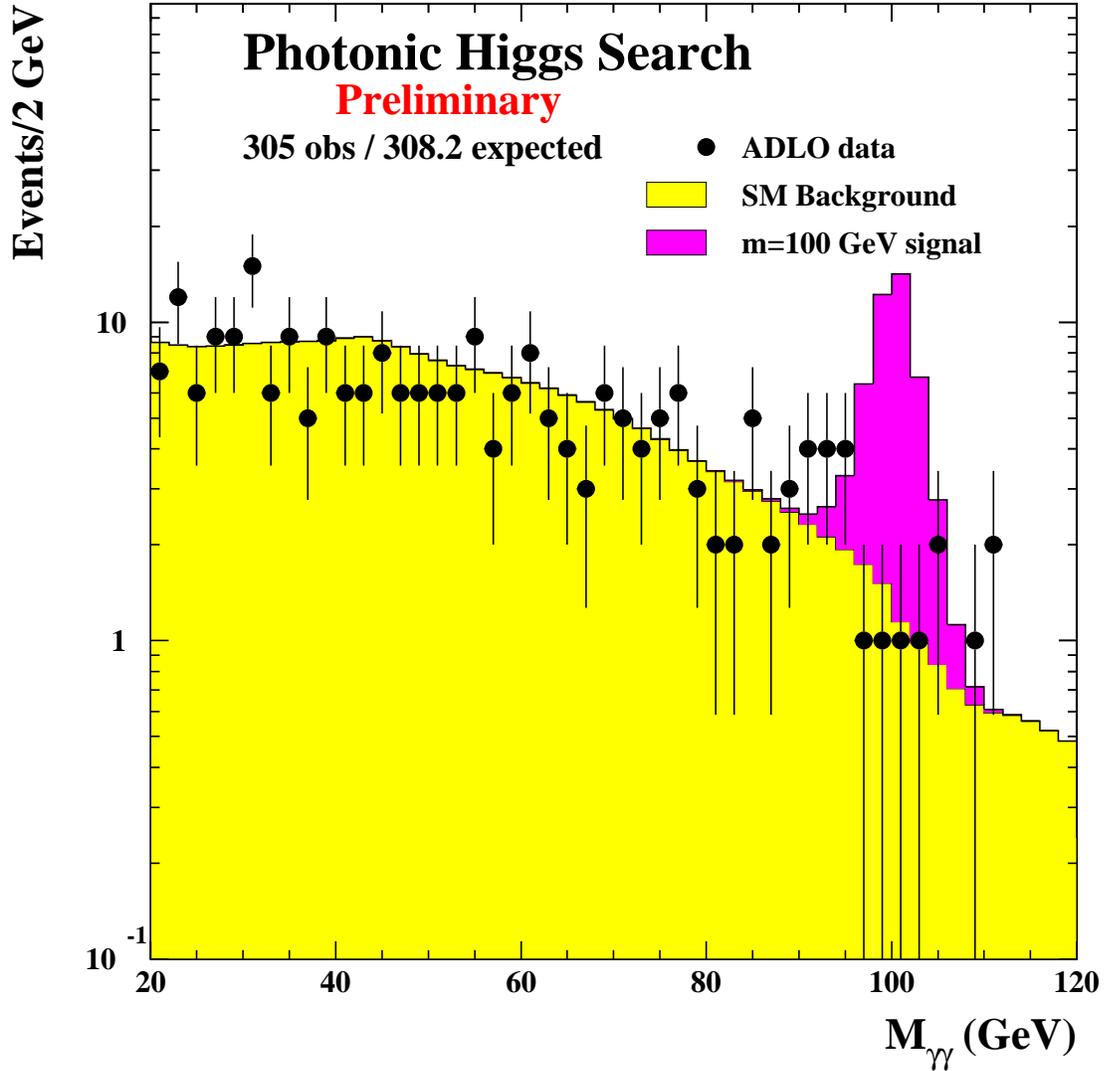} }
        \caption[hgg-mgg]{    
             Distribution of di-photon masses for the LEP experiments
        combined.  The expected background from all SM sources is indicated
        by the lightly shaded histogram.  
        The signal expected for a
        benchmark fermiophobic Higgs boson of mass 100 GeV
        is shown by the dark shaded histogram.
        \label{hgg-mgg} }
        \end{center}
    \end{figure}

The candidates passing the selection cuts for all three decay topologies
are shown in Table~\ref{T:hgg1};
their rate is consistent with the background calculated from SM
physics generators. 
The selected events are used to set an upper limit
on the di-photon branching ratio of a Higgs particle having SM
production rate.  
Figure~\ref{hgg-mgg} showns the di-photon mass for all candidate
events. The distribution appears to be well modelled by the SM background processes.

\begin{figure}[!htb]
        \vspace{0.8cm}
        \begin{center}
           \resizebox{\linewidth}{!}{\includegraphics{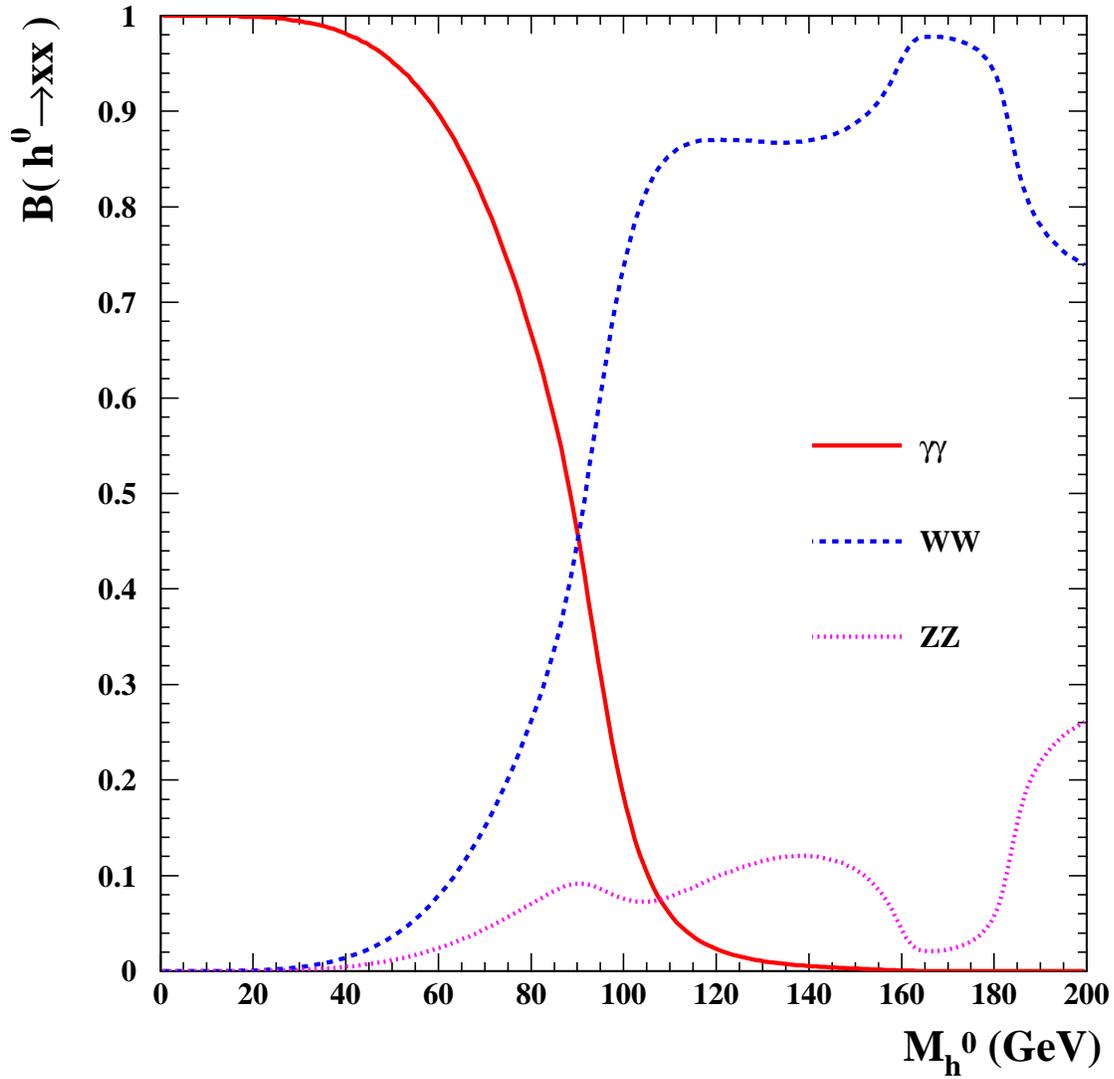} }
        \caption[hgg-br]{    
             Branching fraction of benchmark fermiophobic Higgs boson
        into boson pairs.
        \label{hgg-br} }
        \end{center}
    \end{figure}

\begin{figure}[!htb]
        \vspace{0.8cm}
        \begin{center}
           \resizebox{\linewidth}{!}{\includegraphics{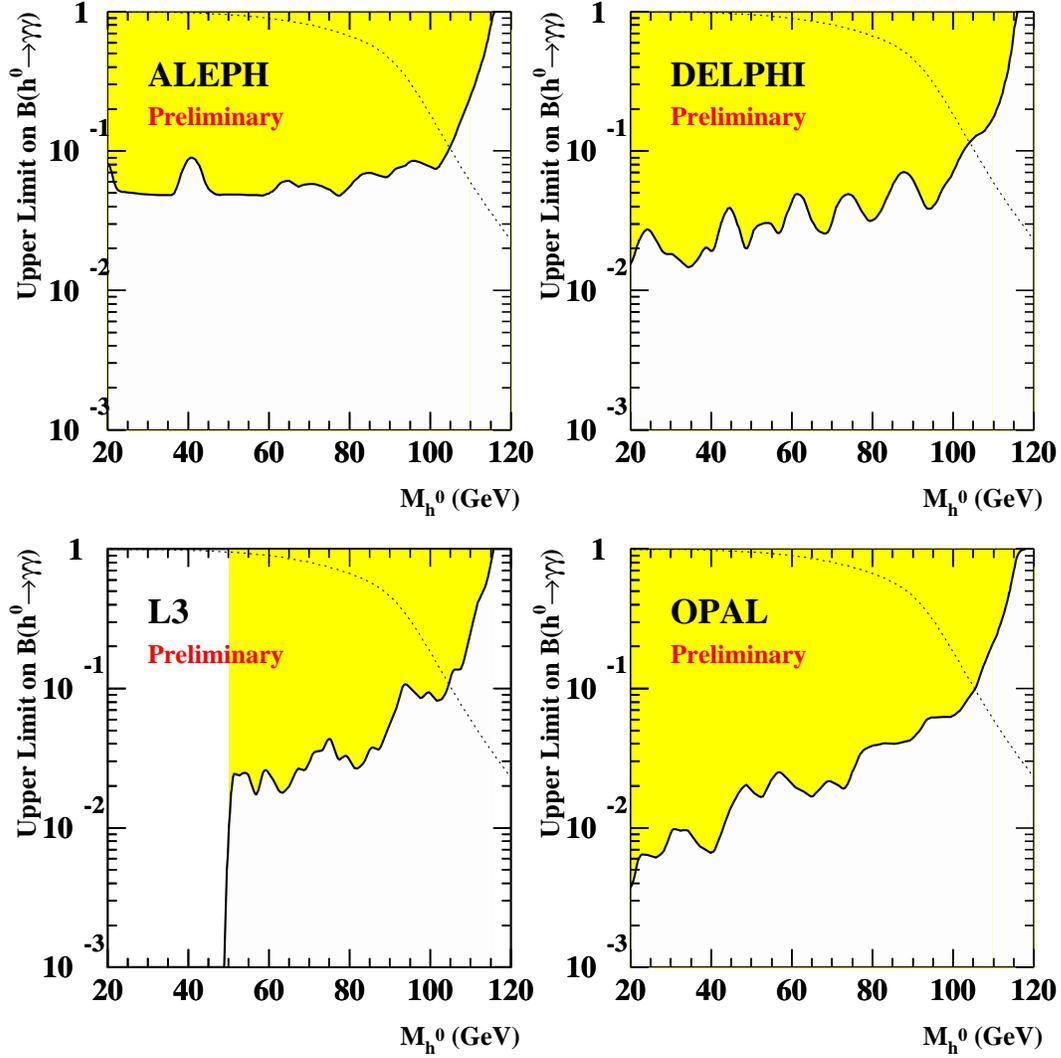} }
        \caption[hgg-adlo]{    
             Individual experimental limits on fermiophobic Higgs
        bosons for the photonic decay mode. The 95\% confidence level upper limit on the di-photon
        branching fraction is shown as a function of Higgs mass;
        the dark shaded regions are excluded. 
        Also
        shown (dotted line) is the branching fraction obtained for the
        benchmark fermiophobic model.

        \label{hgg-adlo} }
        \end{center}
    \end{figure}

\begin{figure}[!htb]
        \vspace{0.8cm}
        \begin{center}
           \resizebox{\linewidth}{!}{\includegraphics{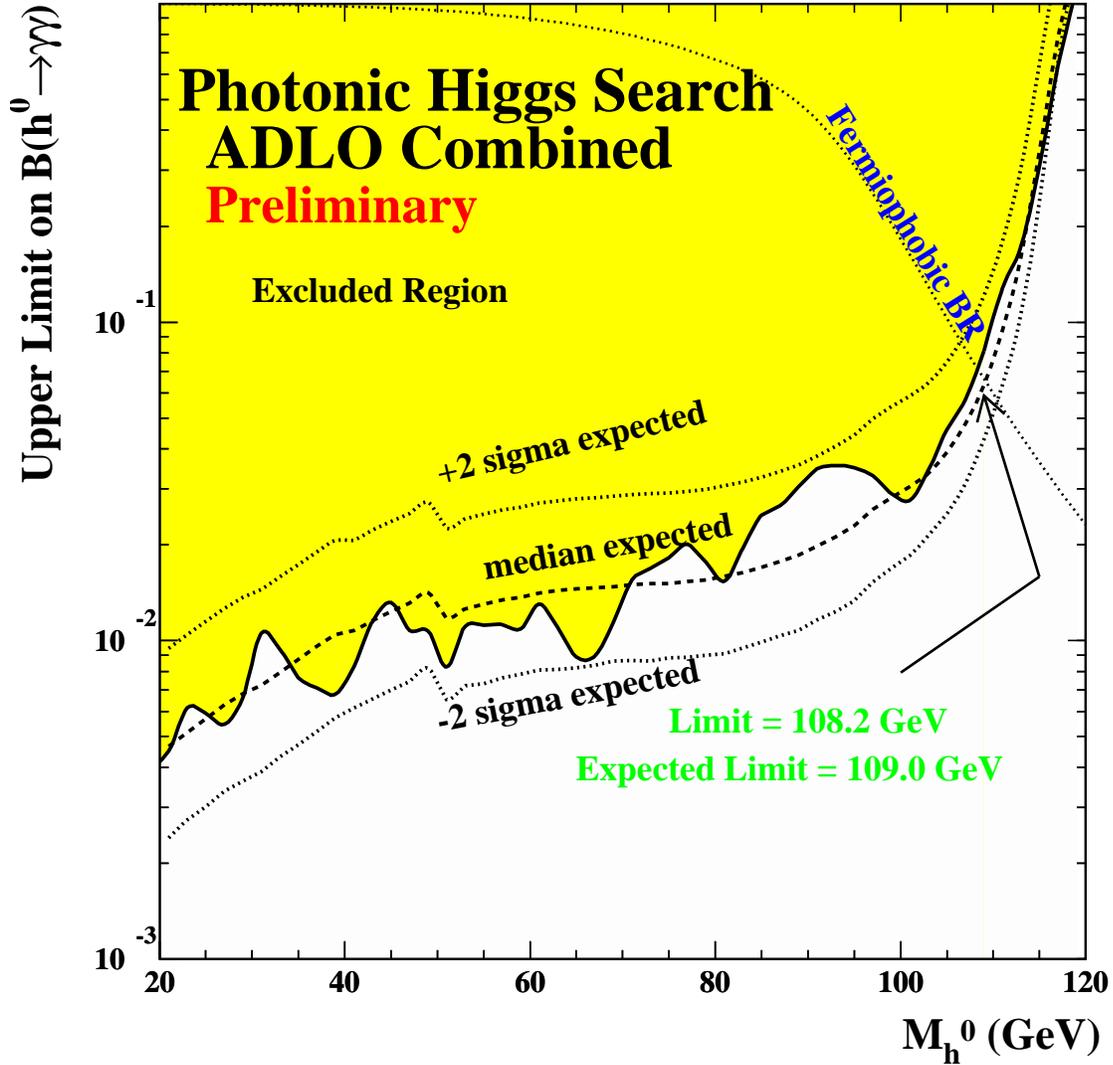} }
        \caption[hgg-s95]{    
             Combined LEP experimental limits on Higgs
        bosons (of Standard Model production cross section)
        decaying into di-photons. 
        The 95\% confidence level upper limit on the di-photon
        branching fraction is shown as a function of Higgs mass. Also
        shown (dotted line) is the branching fraction obtained for the
        benchmark fermiophobic model. The median expected limits and
        the $\pm\sigma$ confidence level region are denoted by the dashed curves.
        \label{hgg-s95} }
        \end{center}
    \end{figure}

Figures~\ref{hgg-adlo} and~\ref{hgg-s95} show the 95\% CL upper limit for the
di-photon branching ratio obtained by combining the candidate events,
where the SM $\ho \mrm Z^{0(*)}$ production cross-section
is assumed at each centre-of-mass energy.
Also shown in the Figures (and in Figure~\ref{hgg-br}) is the $\ho \ra \gaga$ branching
ratio in the Standard Model computed using HDECAY~\cite{HDECAY} 
with the fermionic couplings switched off.
Note that the limits only include the ALEPH data for $\sqrts \ge$ 192~GeV, 
and the L3 inputs to the LHWG only cover the mass range 50-120~GeV.
A 95\% CL lower mass limit 
for such benchmark fermiophobic Higgs bosons 
is set at 108.2~GeV, where the predicted branching ratio 
crosses the upper-limit curve.  The median limit one would expect to obtain in
an ensemble of experiments in the absence of a signal is 109.0~GeV;
if HZHA is used instead of HDECAY to calculate the Higgs boson
branching fractions, the lower mass limits are raised by 0.6~GeV.


%
\clearpage
\newpage

\clearpage
\newpage

\clearpage
\newpage

\clearpage
\newpage

\end{document}